\documentclass[a4paper,11pt]{article}
\usepackage{siunitx}
\usepackage{ulem}

\usepackage{jinstpub} % for details on the use of the package, please
\usepackage{booktabs}
\usepackage{jinstpub}
\usepackage{lineno}
\usepackage{graphicx}% Include figure files
\usepackage{dcolumn}% Align table columns on decimal point
\usepackage{bm}% bold math
\usepackage{hyperref}% add hypertext capabilities
\title{\boldmath Characterization of CRYO ASIC for charge readout in the nEXO experiment}
\author[1]{Z.~Li}% Hawaii, - opted-in
\author[2]{M.~Yu} %SLAC - opted-in
\author[3]{E.~Angelico} % Stanford  opted-in
\author[4]{A.~Atencio} %Drexel - opted-in
\author[2]{A.~Gupta} %SLAC - opted-in
\author[5]{P.~Knauss} %UCSD - opted-in
\author[2]{A.~Pena-Perez} %SLAC - opted-in
\author[2]{B.~G.~Lenardo} %SLAC - opted-in
\author[6]{P.~Acharya} %Alabama - opted-in
\author[7,8]{A.~Amy} % SUBATECH, UMass -opted-int
% ...
\affiliation[1]{Department of Physics and Astronomy, University of Hawaii at Manoa, Honolulu, HI 96822, USA}
\affiliation[2]{SLAC National Accelerator Laboratory, Menlo Park, CA 94025, USA}
\affiliation[3]{Physics Department, Stanford University, Stanford, CA 94305, USA}
\affiliation[4]{Department of Physics, Drexel University, Philadelphia, PA 19104, USA}
\affiliation[5]{Physics Department, University of California San Diego, La Jolla, CA 92093, USA}
\affiliation[6]{Department of Physics and Astronomy, University of Alabama, Tuscaloosa, AL 35405, USA}
\affiliation[7]{SUBATECH, Nantes Universit\'e, IMT Atlantique, CNRS/IN2P3, Nantes 44307, France}
\affiliation[8]{Amherst Center for Fundamental Interactions and Physics Department, University of Massachusetts, Amherst, MA 01003, USA}
% ...

\abstract{nEXO is a proposed next-generation experiment searching for the neutrinoless double beta decay of $^{136}$Xe using a tonne-scale liquid xenon (LXe) time projection chamber (TPC). To image the ionization signals from events in the LXe, the detector will employ metallized fused-silica charge collection tiles instrumented with cryogenic application-specific integrated circuits (ASICs), referred to as CRYO ASIC. These ASICs are designed to operate directly in LXe to minimize input capacitance and pick-up noise. Here we present the performance of the CRYO ASIC mounted on an auxiliary printed circuit board and evaluated in both a cryogenic environmental chamber and a dedicated LXe test stand. 
In the environmental chamber we demonstrate that the ASICs meet nEXO’s gain stability requirement at LXe temperatures, with relative response variations below 0.2\% for nearly all channels over a measurement period exceeding 14 hours. Consistent results were obtained using both the on-chip pulser and an external reference signal. In the LXe test stand, we show that boiling caused by the chip heat dissipation can be mitigated by operating the system above $\sim0.1$ MPa. The in-LXe noise measurement agrees with simulation, which indicates that the $150 ~e^{-}$ design requirement can be satisfied.  These results establish CRYO ASIC as a viable low-noise in-LXe charge readout solution for nEXO and other LXe-based detectors.}

\begin{document}
\maketitle
\flushbottom

\section{Introduction} \label{Intro}

The nEXO experiment is a proposed next-generation search for neutrinoless double beta decay ($0\nu\beta\beta$), aiming to achieve a half-life sensitivity beyond $10^{28}$~years using an isotopically-enriched liquid xenon (LXe) time projection chamber (TPC)~\cite{nEXO:2021ujk}. One of the most important aspects of the TPC is its ionization readout system, which provides both energy information from the integrated charge signal collected at the anode and three-dimensional position and event topology information using the ($x$, $y$) position of the collected charge and the ($z$) position inferred from the time profile of the ionization signals relative to a prompt scintillation signal. 
This information enables powerful background rejection: $0\nu\beta\beta$ decays would produce a highly-localized monoenergetic signal at $Q_{\beta\beta} = 2.457\,\,\mathrm{MeV}$, corresponding to approximately $10^5$ ionized electrons, while background gamma rays from the decay of trace $^{214}$Bi and $^{208}$Tl in detector construction materials produce energy spectra characteristic of their respective $\gamma/\beta$ emissions and can deposit their energy at multiple locations via Compton scattering.
The design of the nEXO charge readout system is driven by three primary needs: 1) maintaining extremely low radioactivity, 2) achieving mm-scale position reconstruction for background discrimination and 3) achieving an overall energy resolution better than 1\% at the $\beta\beta$ Q-value of 2.458\, MeV. These are addressed in the nEXO conceptual design via a system consisting of an array of ultra-low-background sensor tiles with crossed strips with on-board readout electronics. This system requires a readout solution capable of 1) digitizing 3,580 independent channels with 2) microsecond-scale time resolution while 3) maintaining electronic noise below 150 e$^-$ equivalent noise charge (ENC) and 4) allowing precise cross-calibration between channels for complete event-level charge reconstruction.

\begin{figure}
   \begin{minipage}[t]{0.45\textwidth}
    \vspace{0pt} % force top alignment
    \centering
    \includegraphics[width=\linewidth]{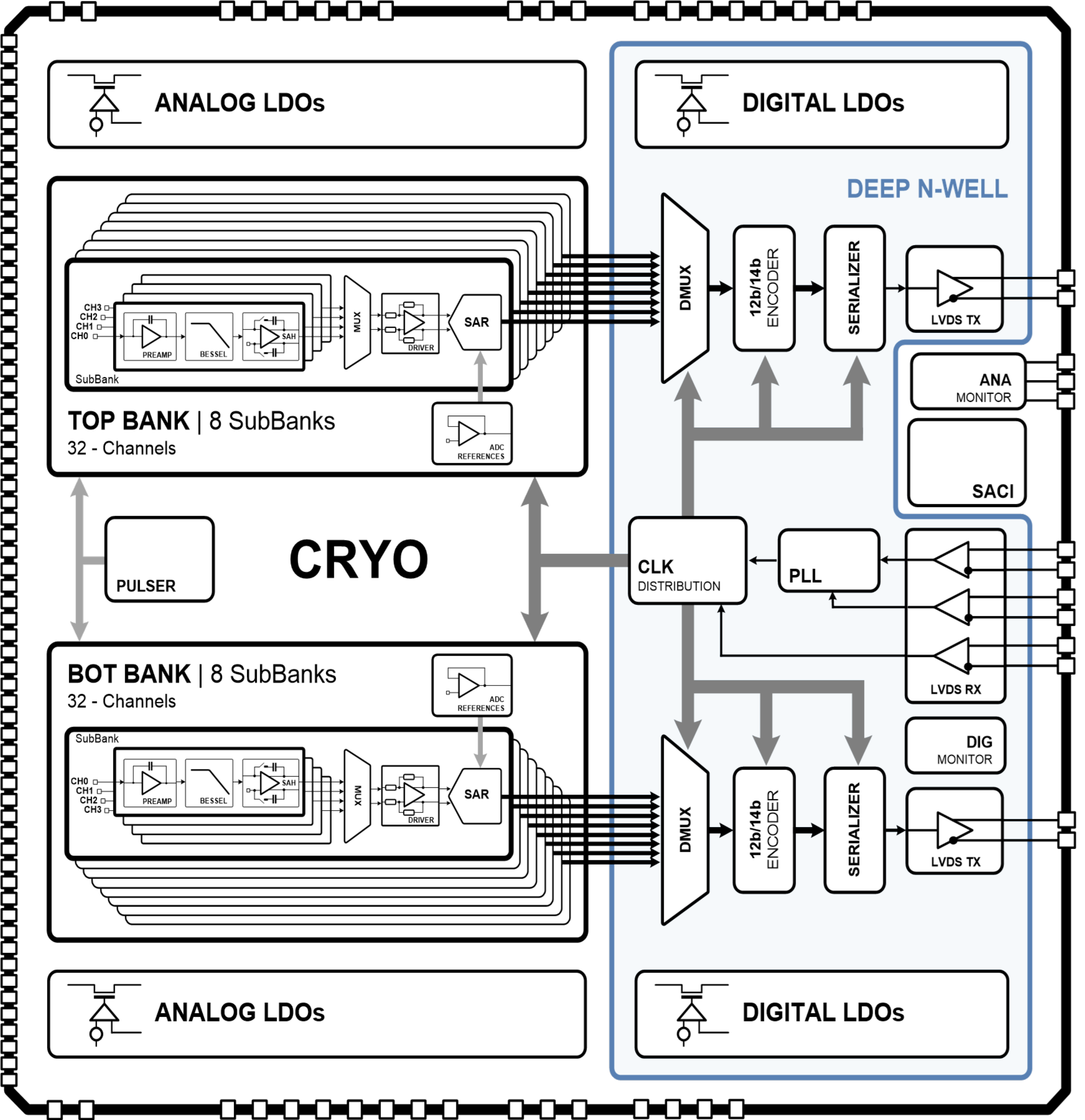}
   \end{minipage}\hfill
   \begin{minipage}[t]{0.45\textwidth}
    \vspace{0pt} % force top alignment
    \centering
    \begin{tabular}{@{}ll@{}}
      \toprule
      \textbf{Parameter} & \textbf{Specification} \\
      \midrule
      Process           & 130~nm CMOS \\
      Supply voltage    & 2.5~V (2~V, 1~V internal) \\
      Input capacitance & $\sim$~20~pF -- 30~pF \\
      Anti-aliasing filter & 5th order Bessel architecture \\
      Peaking times     & 0.6~$\mu$s, 1.2~$\mu$s, 2.4~$\mu$s, 3.6~$\mu$s \\
      Gain settings     & 9.6~mV/fC, 14.3~mV/fC, \\
                        &28.6~mV/fC, 57.2~mV/fC \\
      Max. input charge & 150~fC, 100~fC, 50~fC, 25~fC \\
      Noise             & $< 150\,e^{-}$ @28.6 mV/fC and 1.2 $\mu s$ \\
      ADC               & 12-bit 2~MSPS / CH \\
      INL and DNL       & $\pm$1 LSB \\
      Power consumption & $< 20$~mW/CH  \\
      Temperature       & LXe $\sim$165~K (-108~\textdegree{}C) \\
      \bottomrule
    \end{tabular}
   \end{minipage}
   \caption{Functional block diagram of the CRYO ASIC and the design specifications of the nEXO-specific implementation.} %The power consumption is specified for a future 32-channel version of the ASIC.}
   \label{fig:block_diagram}
\end{figure}

To meet these goals, the nEXO collaboration is pursuing a custom cryogenic application-specific integrated circuit (ASIC), referred to as CRYO ASIC, for in-LXe charge readout. CRYO ASIC is a System-on-Chip (SoC) waveform digitizer and serializer, implemented in TSMC's 130~nm CMOS process, designed specifically for noble liquid TPC experiments~\cite{9507812, Pena-Perez:2022zvt,Doering:2022ayx}. The functional block diagram of the ASIC is illustrated in Fig.~\ref{fig:block_diagram} alongside the key specifications for the nEXO experiment. The ASIC integrates both analog and digital functions into a single chip, providing 64 input channels, each sampled at 2~MSPS with 12-bit resolution across a 1.6~V range. The digitized data are output at a bit rate of 896~Mbps.The gain and peaking time of the front end are each configurable between four different values to optimize between resolution, dynamic range, and noise. 
To comply with the stringent low-radioactivity requirements, the design minimizes the need for external components by employing on-chip supply regulation with low-dropout (LDO) voltage regulators distributed across the chip.
CRYO ASIC has been designed using known best-practices for extended operation at cryogenic temperatures~\cite{Hoff:2015hax}, and enables the user to correct for temperature-dependent effects by directly programming the voltage levels in core sections through internal registers. The ASIC and data acquisition (DAQ) communicate via SACI, a custom serial communication interface developed at SLAC National Accelerator Laboratory~\cite{slac_saci}. 
An internal pulser system is integrated to facilitate \textit{in-situ} testing of the front-end channel performance. The pulser comprises a single shared 10-bit programmable DAC which routes to individual test capacitors and switches at the front end of each channel. The enable/disable state and the amplitude of the internal pulser are configurable through internal registers accessed via the SACI interface. The differential nonlinearity and integral nonlinearity of the system are both~0.5~LSB.

In this paper, we present new results from cryogenic testing and characterization of the first prototype CRYO ASIC for nEXO. We demonstrate the basic functionality and stability of the ASIC at 165~K in a dry nitrogen (N$_2$) gas environment, then report successful operation in LXe that meets nEXO's noise requirements. A key result is the demonstration of the mitigation of LXe boiling to enable stable in-LXe operation. This paper is organized as follows: in Section~\ref{Sec:Exp} we describe the ASIC DAQ systems and the cryogenic test platforms used in this work, then Section~\ref{Sec:Results} describes our measurement results. We conclude with a discussion of the context of these results within the nEXO project and needs for future testing.

\section{Experimental Apparatus}~\label{Sec:Exp}
%UCSD setups (Zepeng)
%The nEXO experiment's strict radiopurity and mechanical requirements will necessitate extreme control of materials used for external components and cabling. 
For testing purposes, we use commercially available printed circuit boards~(PCBs) with off-the-shelf discrete components, and use loading capacitors to characterize the front-end performance under realistic conditions. These test boards and test stands are described in detail here.

\subsection{ASIC test setups}

\begin{figure}
    \centering
    \includegraphics[width=0.95\linewidth]{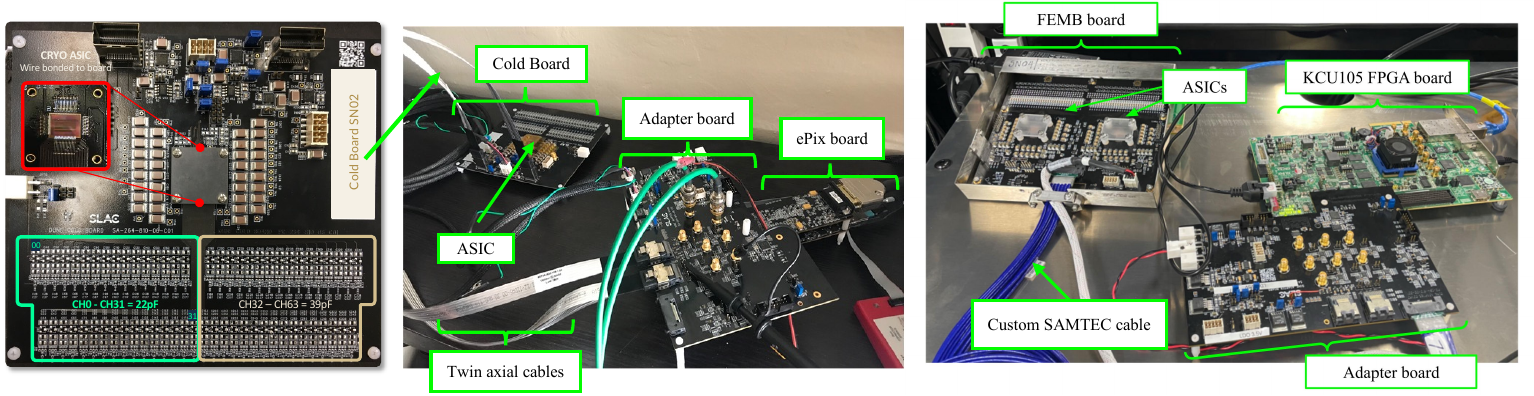}
    \caption{Photos of the CRYO ASIC testing setups. Left: The CRYO ASIC mounted on the Cold Board, a custom PCB for single-chip testing. The inset shows the ASIC wire bonded to the board, which is then hidden below a protective cover during handling.  The 64 input channels are loaded with 22~\text{pF} or 39~\text{pF} capacitors. Middle: the cold board communicates with the ePix FPGA board for data acquisition and uses an adapter board for power management and signal routing. Right: The CRYO ASIC FEMB Board mounted in an RF-shielded box, with the two CRYO ASICs protected beneath acrylic covers. It interfaces with the adapter board and is controlled via a commercial KCU105 FPGA board. }
    \label{fig:ASIC_testing_boards}
\end{figure}

We use two separate systems to communicate with and test the ASICs. The first is the CRYO ASIC ``Cold Board,'' a compact PCB designed to host a single CRYO ASIC chip and enable loading of the ASIC inputs to simulate the detector properties. The left panel of Fig.~\ref{fig:ASIC_testing_boards} shows the Cold Board with one CRYO ASIC directly mounted and wire-bonded to the board. 
The use of bare dies is adopted for radiopurity reasons: conventional ASIC packaging contains up to $10^4$ ppt of U/Th contamination whereas bare dies have reported impurity concentrations below 1~ppt~\cite{MacLellan:2026tik}. 
The Cold Board supports two power schemes: (1) providing the main 2.5~V supply via an internal LDO regulator and allowing the on-chip circuitry to distribute power, and (2) disabling the on-chip LDOs and using external LDOs to supply the required 2~V and 1~V to the analog and digital circuitry, respectively. 
In this work we use option 1. We load the 64 input channels of the CRYO ASIC with \(22~\text{pF} \, (\pm 1\%)\) or \(39~\text{pF} \, (\pm 1\%)\) ceramic capacitors to study the noise performance as a function of the input capacitance. We choose C0G capacitors, whose values change by less than 0.3\% when cooled down to LXe temperature ($\sim$165~K).

The programming and testing of the Cold Board 
is accomplished using a custom Field Programmable Gate Array (FPGA) board developed for the ePix camera system~\cite{nishimura2016design}, shown in the middle panel of Fig.~\ref{fig:ASIC_testing_boards}.
The ePix board interfaces with an adapter board to supply power to the Cold Board and enable communication between the Linux desktop computer and the Cold Board. Data and control signals are transmitted using two high-speed twin axial cables~\cite{twinax}. 
The register control and programming of the ASIC are managed through the SACI protocol~\cite{slac_saci}. The Linux desktop is configured with the open-source SLAC SURF~\cite{surf} and ROGUE~\cite{rogue} environments to support data acquisition.

The second CRYO ASIC test board, shown in the rightmost panel of Fig.~\ref{fig:ASIC_testing_boards}, is the Front-End Mother Board (FEMB), which is larger and allows the simultaneous operation of two ASICs.
Like in the Cold Board system, we operate the ASICs with their internal supply regulation and the FEMB board supplies the main 2.5~V. Unlike the Cold Board, where the input capacitors are placed directly between ground and each channel input, the FEMB routes the ASIC inputs to connectors on the backside of the board, allowing convenient connection to different input loads. 
In addition, the FEMB system uses a commercial KCU105 FPGA board~\cite{kcu105} in place of the custom ePix board, improving portability across different test platforms. The same adapter board used in the Cold Board system is reconfigured to enable communication and control through the commercial FPGA. Data are transmitted using a custom high-speed cable from SAMTEC. 

During cryogenic testing, the CRYO ASIC test board (Cold Board or FEMB) is operated at cryogenic temperatures, while the remaining DAQ electronics are maintained at room temperature.

\subsection{Cryogenic test stands}

\begin{figure}
    \centering
    \includegraphics[width=0.95\linewidth]{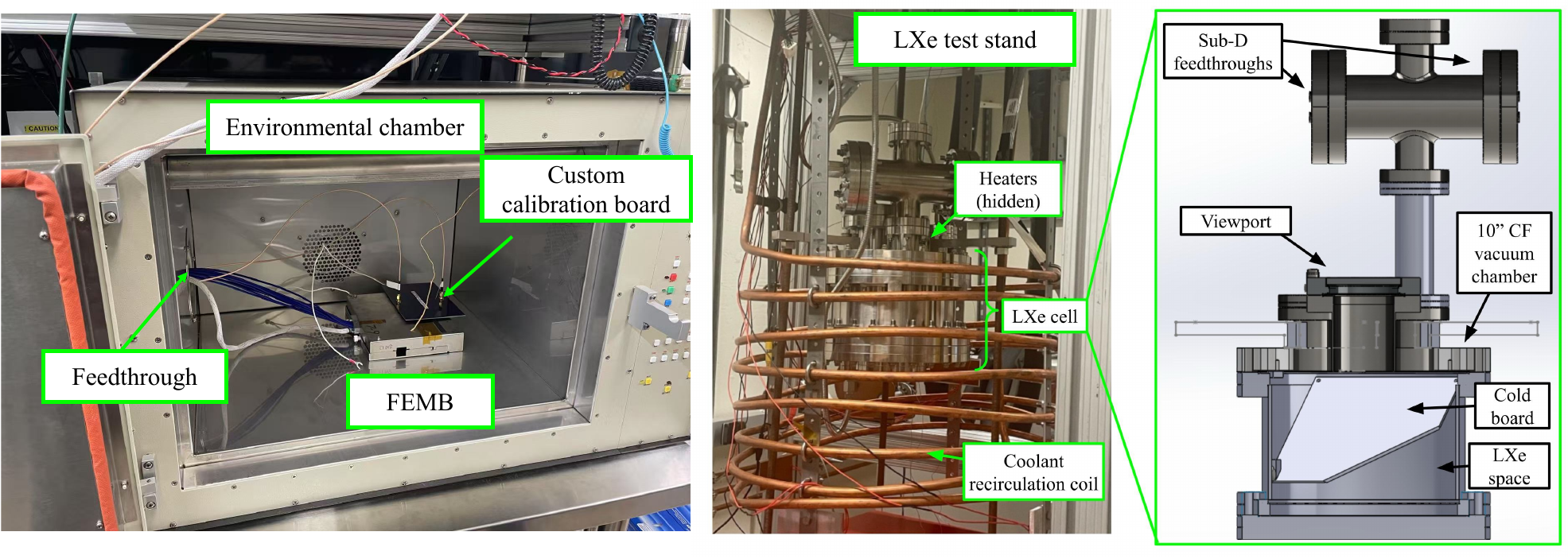}
    \caption{Left: A photo of the liquid nitrogen-cooled environmental chamber. The FEMB board is placed in the chamber connected to the DAQ system at room temperature. A custom calibration board (purple), plugged into the back of the FEMB board for gain calibration studies, can also be seen in the photo. Middle: Photo of the inner LXe chamber installed inside the outer cryostat. Right: schematic drawing of the inner LXe chamber and placement of the cold board inside the chamber. }
    \label{fig:cryogenic_test_stands}
\end{figure}

A liquid-nitrogen-cooled environmental chamber~\cite{sunelectronicsEC02Temperature}, shown in the left panel of  Fig.~\ref{fig:cryogenic_test_stands},  is used to test ASIC boards in a dry N$_2$ gas environment at the LXe temperature of 165~K. The chamber employs liquid nitrogen evaporation for cooling and trim heating for temperature regulation. A fan continously recirculates the cold N$_2$ gas at a flow rate of \SI{4.25}{m^3/min} to maintain temperature uniformity throughout the chamber. The system also allows adjustable temperature ramp rates, enabling rapid and precise control of the chamber temperature. The metal enclosure of the chamber provides shielding from environmental radio-frequency (RF) noise, particularly electromagnetic radiation at radio frequencies. 
%The feedthroughs are routed into the chamber through the side-wall port. 
Cables are routed into the chamber through feedthroughs in the side-wall port.
The environmental chamber offers a flexible and easily configurable testing environment, with a large internal volume that accommodates boards and setups of different sizes.

In addition to the N$_2$ environmental chamber, to replicate the nEXO operating environment we use a dedicated test stand for in-LXe measurements at the University of California San Diego. The test stand consists of a xenon gas handling system, an inner LXe chamber, and an outer HFE-7000 cryostat, similar to the systems described in Ref.~\cite{Angelico:2025cpz}. The system is shown in Fig.~\ref{fig:cryogenic_test_stands}; the middle panel shows a photograph of the LXe chamber installed inside the outer cryostat, while the right panel of Fig.~\ref{fig:cryogenic_test_stands} presents a schematic drawing of the LXe chamber. The inner LXe chamber is constructed from a \SI{25.4}{cm} ConFlat spool-piece and two flanges. The chamber body is made of 304 stainless steel, with an inner diameter of 20~cm and a height of 12~cm. The top flange includes a gas feedthrough for xenon filling and recovery, a ConFlat 4-way reducing cross for electrical feedthroughs, and a ConFlat quartz viewport. The 4-way reducing cross has two 50-pin ~(25-pin$~\times$~2) D-type feedthroughs, enabling the connection of four custom cables terminated with D-type connectors that link the Cold Board to the adapter board. An Arducam Motorized Focus Pi OV5647 Camera~\cite{arducamArducamAuto} is mounted above the viewport to visually monitor the CRYO ASIC and LXe conditions. The camera is controlled and read by a Raspberry Pi located outside the outer cryostat. The bottom flange includes a second gas feedthrough for xenon filling and recovery. During operation, the inner chamber is filled with $\sim$7 kg of LXe to fully submerge the CRYO ASIC mounted on the Cold Board. Fig.~\ref{fig:monitor-camera} shows images of the Cold Board immersed in LXe, captured by the monitoring camera through the viewport. 
%\begin{figure}
%    \centering
%    \includegraphics[width=0.5\linewidth]{figures/Figure4.pdf}
%    \caption{ Image of the Cold Board immersed in LXe, %captured by the monitoring camera through the viewport. The halo observed near the center of the image is caused by light from LEDs surrounding the camera reflecting off the LXe surface. }
%%    \label{fig:camera_lxe}
%\end{figure}

The test stand is typically operated at $\sim$168~K and at pressures below 0.15~MPa. Following the cryogenic scheme used in EXO-200 and proposed for nEXO~\cite{EXO-200:2021srn}, the LXe chamber is submerged in HFE-7000 cooling fluid to maintain good temperature uniformity. The HFE-7000 is contained within a cryostat and cooled by a large copper coil immersed in the fluid, which recirculates coolant from a Telemark Model 1800 cryochiller. Four thermocouples monitor the system temperature: three are attached at different locations on the inner chamber, and one is positioned at the bottom of the HFE-7000 bath in the outer cryostat. A PID control loop, implemented in LabVIEW, regulates the coolant flow based on measurements from the thermocouples mounted in the middle of the LXe chamber, thereby stabilizing the system temperature.

\section{Results}~\label{Sec:Results}
Simulations of ionization signals in nEXO identify a gain of 28.6 mV/fC as the preferred configuration: this setting maximizes signal-to-noise while keeping signals at $Q_{\beta\beta}$ well within the dynamic range of the ASIC. All of the following measurements are conducted at this gain setting.

\subsection{CRYO ASIC functionality tests in cryogenic environmental chamber}

The initial performance of CRYO ASIC is evaluated in the dry N$_2$ environmental chamber with both Cold Board and the FEMB boards, as illustrated in the left panel of Fig~\ref{fig:cryogenic_test_stands}. Environmental chamber tests of the Cold Board primarily focused on register-level optimization across temperatures, while the FEMB board (which features a more configurable input load) is used for more comprehensive ASIC characterization studies, including the pulse response, noise performance, and gain stability measurements. Both boards are placed inside metal boxes for RF shielding and a temperature sensor is installed inside the shielding enclosure during testing to directly monitor the temperature at the ASICs. 

% Tests of the Cold Board primarily focus on register-level optimization across temperatures ranging from room temperature down to 165~K. The optimized register settings for the LXe temperature are then applied to all tests presented in this work. By taking advantage of customizing the channel inputs through connectors, comprehensive ASIC characterization studies, including the pulse response, noise performance with different configurations and gain stability measurement, are conducted with the FEMB board. 
%A photograph of the FEMB board installed in the environmental chamber can be seen in the left panel of Fig.~\ref{fig:cryogenic_test_stands}.

\begin{figure}
    \centering
    \includegraphics[width=0.9\linewidth]{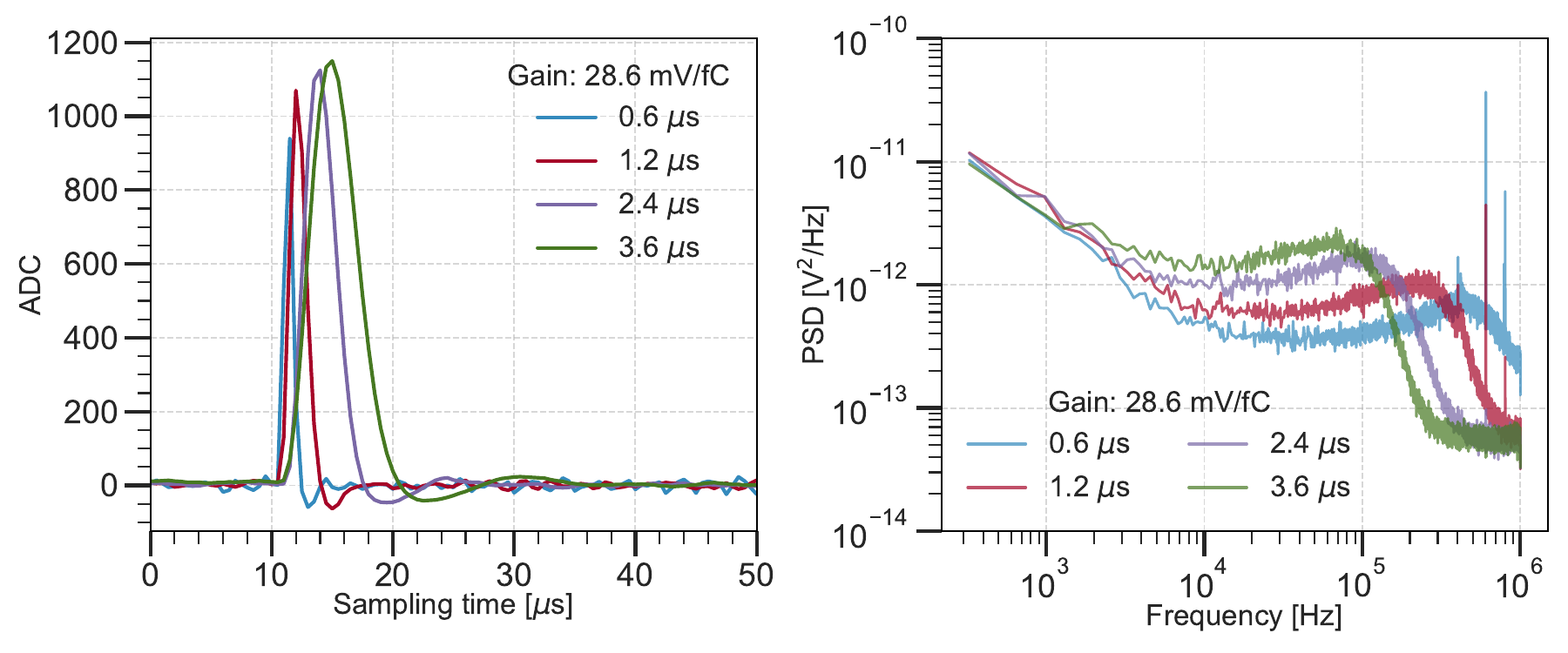}
    \caption{ASIC characterization in the nitrogen environmental chamber. Left: sample waveforms from FEMB board with internal pulses. Right: corresponding noise power spectral density of four peaking time settings with 28.6 mV/fC gain.}
    \label{fig:waveforms_and_PSD_FEMB}
\end{figure}

The initial calibration procedure is done by adjusting the ASIC's internal registers so that the voltage levels of the internal LDOs, verified through analog and digital monitor outputs, match the expected design values (2~V and 1~V, respectively). This corrects for changes in the behavior of the 130~nm CMOS devices as a function of temperature. The calibration procedure is performed at each operating temperature (in this work, room temperature and 165~K) and provides a table of optimized register configuration values which are then loaded during the ASIC initialization process before making measurements at a given temperature.

Next, we use the FEMB system to verify the pulse response at the four different peaking times using the ASIC's internal pulser. Data acquisition is trigger by the internal pulse, therefore, pulse peaks occur at a consistent position within the acquisition window, as shown in the left panel of Fig.~\ref{fig:waveforms_and_PSD_FEMB}. 
To evaluate the noise performance, the internal pulser is disabled, and the data collection is then randomly triggered. 
A total of 50 traces are recorded sequentially, with 6000 ADC samples collected per channel in each trace. The power spectral density (PSD), shown in the right panel of Fig.~\ref{fig:waveforms_and_PSD_FEMB}, is computed using a discrete-time Fourier transform for each trace and subsequently averaged.
The characteristics of the anti-aliasing filter
are clearly observed in the spectra: configurations with larger peaking times (i.e., smaller bandwidth) more effectively suppress high-frequency components. Narrow spikes at high frequencies are present during testing, coming from discrete interference tones picked up from the surrounding environment. These features do not contribute significantly to the integrated noise.

With the ASIC's operation and noise spectra verified, we then study stability over time. Achieving nEXO’s target energy resolution (<1\% at the $\beta\beta$ Q-value) requires calibrating the gain of each channel to within 0.2\%, with periodic recalibrations throughout the planned 10-year operation. The regular monitoring can be realized with the internal pulser system, which tracks the charge response of all front-end channels over time. To characterize the stability of the internal pulser, an external calibration system is implemented to inject charge directly into the front-end channel inputs. A custom calibration board, connected to the SAMTEC connector of the FEMB board as shown in the left panel of Fig.~\ref{fig:cryogenic_test_stands}, is designed to simultaneously inject charge into 64 channels. For each channel, a 22~pF capacitor is connected between the channel input and the ground to simulate the detector capacitance. A 1~\text{pF} capacitor is connected in series between the channel input and an external pulser to convert voltage pulses into charge signals. The external pulser generates a square waveform with an amplitude of 40 $\mathrm{mV_{pp}}$ and at a frequency of 3~kHz.
All capacitors are C0G type to minimize temperature-dependent variations, and the traces for all channels on the calibration board are designed to ensure nearly uniform contributions from parasitic capacitances, which are estimated to be approximately 2-3~pF.
\begin{figure}
    \centering
    \includegraphics[width=0.95\linewidth]{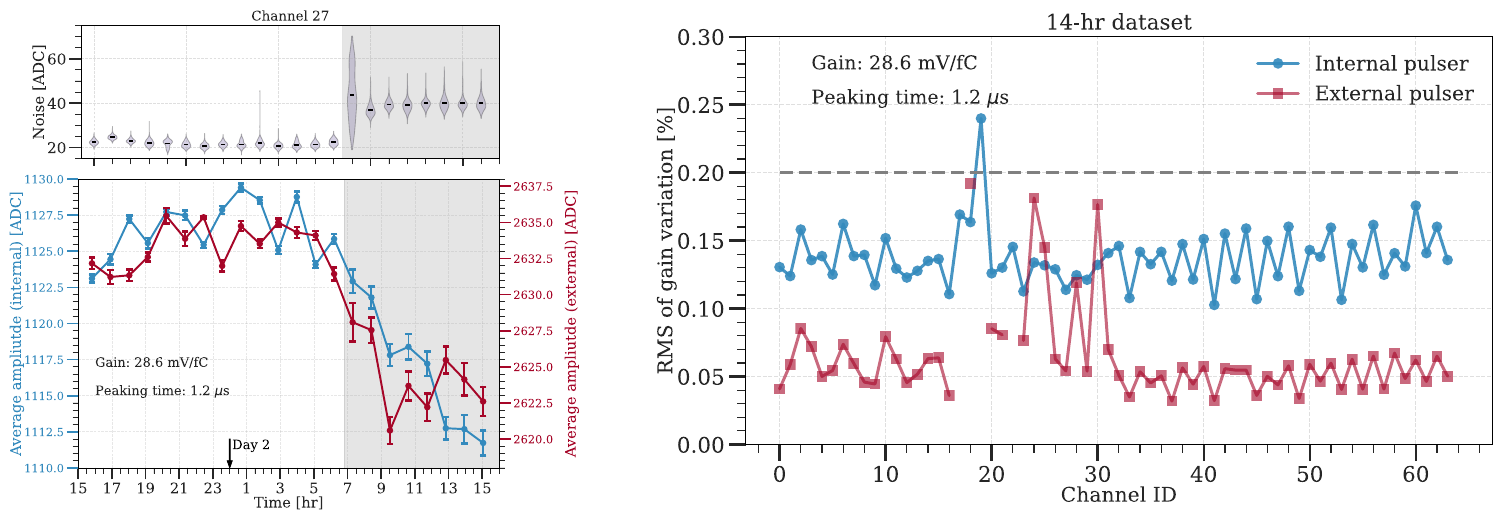}
    %\caption{Relative noise level (upper left) and gain variation (lower left) over 24-hr overnight dataset on one representative channel and the RMS of gain variations of all 64 channels for both internal pulses and external pulses (right). The peaking time is set as 1.2 $\mu$s with a gain of 28.6 mV/fC. During the 14 hours, the noise level of the ASIC remained consistently low. Correspondingly, gains calibrated by both pulsers which track each other well, are stable and basically satisfy the <0.2\% variation requirement for nEXO. No external pulses are found in three channels on the top bank, which results from bad soldering of the capacitors on the charge-injection board. }
    \caption{Upper left: noise variation of one representative channel (channel 27). Lower left: relative gain variation over 24-hr overnight dataset on this channel. Right: RMS of gain variations of all 64 channels for both internal pulses and external pulses. The peaking time is set as 1.2 $\mu$s with a gain of 28.6 mV/fC. During the 14 hours, the noise level of the ASIC remained consistently low. Correspondingly, gains calibrated by both pulsers which track each other well, are stable and basically satisfy the <0.2\% variation requirement for nEXO. No external pulses are found in three channels on the top bank, which results from bad soldering of the capacitors on the charge-injection board.}
%    \caption{Relative variation of the input gain over time, normalized to the last data point, which is measured in the nitrogen environmental chamber. Channel 62 is shown as an example, with the gain set to 28.6 mV/fC and the peaking time to 1.2  $\mu$s. Calibration results obtained using the internal pulser and the external pulser are shown in blue and red, respectively. The RMS gain variations for the internal and external pulser measurement are 0.12\% and 0.10\%, respectively, satisfying the <0.2\% variation requirement for nEXO. The strong correlation between the two measurements indicates that the internal pulser can provide reliable in-situ calibration. }
    \label{fig:gain_calibration}
\end{figure}

The gain is monitored over a 24-hour period in the nitrogen cryogenic environmental chamber at 165~K. The channel gains are measured once per hour using both the internal and the external pulser calibration systems. The external pulse generator and associated cabling add significant noise to the input when connected, and several grounding configurations were tested to minimize pickup noise to acceptable levels ($\sim20$ ADC RMS). Still, we observe significantly ($>2\times$) elevated noise during the 8am -- 4pm working hours, shown as the gray shaded region in the upper left panel of Fig.~\ref{fig:gain_calibration}, which induces a systematic shift in our reconstructed gain compared to lower-noise data. As this noise is only present when the external generator is connected directly to the ASIC inputs, it is not a concern for the nEXO project, and we here restrict our analysis to data taken between the hours of 4pm -- 6am, when noise performance is best.

%The noise level on one representative channel (channel 27) can be found in the upper left panel of Fig.~\ref{fig:gain_calibration}. 
%Approximately 40~fC of charge is injected by either pulser. 
 For each calibration run, approximately 3000 waveforms are recorded and analyzed offline. The pulse amplitude is extracted by fitting each waveform with a Gaussian function, and the average fitted amplitude is used to determine the channel response. The gain variation is defined as the relative change in the average pulse amplitude with respect to its reference value. During these measurements, the front-end channels are configured with a gain of 28.6 mV/fC and a peaking time of 1.2 $\mu$s.
 
 Average amplitudes of both the internal and external pulser signals on one representative channel (channel 27) are shown in the lower left panel of Fig.~\ref{fig:gain_calibration}. The two gain-variation curves track each other well, indicating the ability of self-calibration by the internal pulses.
We observe variation below 0.2\% for all channels in both the internal signals and the external reference signals, as shown in the right panel of Fig.~\ref{fig:gain_calibration}, with the exception of the internal pulser calibration of channel~19. The "bottom" channel bank (channels 32-63) of the measured ASIC exhibits better gain stability than the top bank. The reason for this discrepancy is not yet understood.

Overall, these measurements demonstrate that the prototype ASICs satisfy nEXO's gain stability requirement of 0.2\% over the measured 14-hr period.
Longer-duration measurements under controlled low-noise conditions, together with multi-amplitude gain scans, will be needed to establish calibration accuracy and long-term stability for the nEXO experiment.

\subsection{Study of boiling conditions during LXe operation}\label{subsec:LXeBoiling} 
In the nEXO TPC, heat dissipation from the front-end electronics is a key concern, as it may create local boiling conditions in LXe. Such boiling could generate microphonic noise in the electronics or cabling, or potentially lead to electrical breakdown in regions with high electric field.
Previous work has studied the onset of nucleate boiling in LXe to better understand the relevant boiling conditions~\cite{Breur:2020hzd}. To investigate possible boiling induced by the power dissipation of the CRYO ASIC and to explore the mitigation strategies, the Cold Board is tested in the LXe test stand with the CRYO ASIC immersed in LXe and its protective cover removed. 
The LXe is visually monitored by the camera through the viewport on the top flange of the LXe chamber, and is quantitatively measured using temperature and pressure sensors. Noise measurements can be performed immediately after the camera is turned off to eliminating pickup noise from the camera in the ASIC front-end electronics.

\begin{figure}
    \centering
    \includegraphics[width=0.95\linewidth]{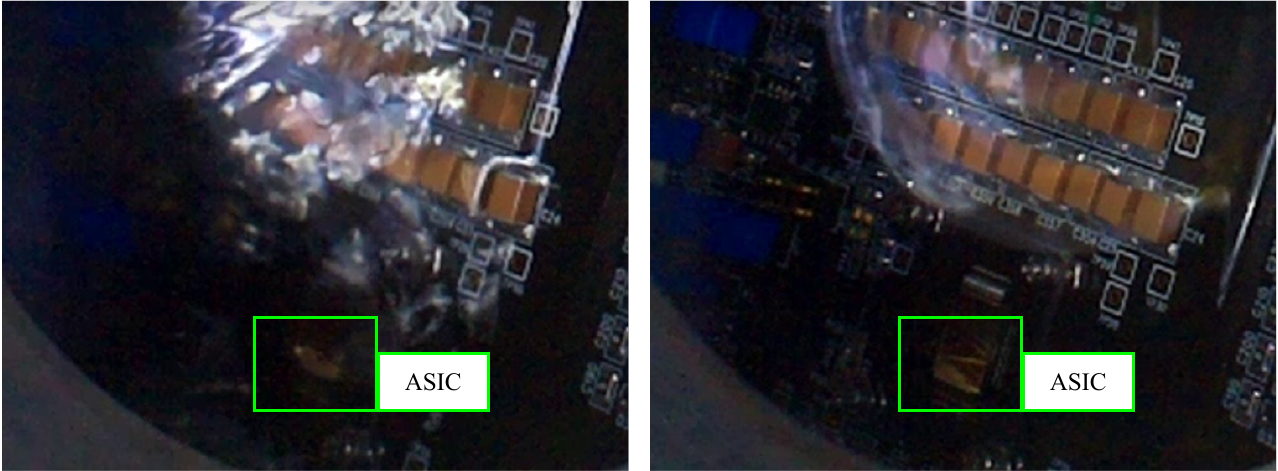}
    \caption{Left: a screenshot of the camera monitoring the LXe condition, where ripples from boiling are clearly visible. The liquid–gas interface is visible near the right side of the image as a faint inclined line crossing several board components
    Right: a screenshot of the camera showing the LXe condition with no visible boiling from the ASIC. The bright patterns on the liquid surface arise from reflections of the camera’s surrounding LED illumination off the LXe surface. }
    \label{fig:monitor-camera}
\end{figure}

Significant boiling is observed after the CRYO ASIC is powered on, once the LXe conditions stabilize at a pressure of 0.095~MPa and a temperature of \SI{170.45}{K}. This is shown in the left panel of Fig.~\ref{fig:monitor-camera}. An increase in noise correlated with the boiling is observed in the measured data, as shown in the noise power spectra of Fig.~\ref{fig:coldboard_LXeTest_PSD}.

To suppress boiling, we increase the xenon gas pressure inside the chamber using two polyimide flexible adhesive heaters mounted on the top flange of the LXe chamber. Because the HFE-7000 coolant level remains below the top flange, the heaters produce only a negligible temperature increase in both the HFE-7000 bath and the inner volume of the LXe chamber and primarily heat the gas phase. We heat the top flange to raise the pressure inside the chamber to 0.104~MPa, which effectively suppresses LXe boiling. The mitigation of LXe boiling is confirmed by both the camera monitoring and noise measurements.
The right panel of Fig.~\ref{fig:monitor-camera} shows a photograph of the LXe recorded by the monitoring camera at the elevated pressure, showing no boiling. 
%Since nEXO is expected to operate at approximately 165~K and pressures around 0.15~MPa~\cite{nEXO:2018ylp}, these results indicate that the CRYO ASIC will not induce boiling under the planned detector operating conditions.
Since nEXO is expected to operate at approximately 165~K and 0.15~MPa~\cite{nEXO:2018ylp}, these results suggest that the planned operating conditions provide a favorable margin against boiling induced by CRYO ASIC heat dissipation.  

\subsection{Noise measurements with CRYO ASIC in LXe} 
The noise performance of CRYO ASIC is studied in the LXe chamber with the system pressure elevated to suppress xenon boiling. Measurements are performed only after the monitoring camera confirms that the ASIC chip is fully submerged in the liquid. During the tests, the system conditions are continuously monitored to ensure that no xenon boiling occurs.

\begin{figure}
    \centering
    \includegraphics[width=0.95\linewidth]{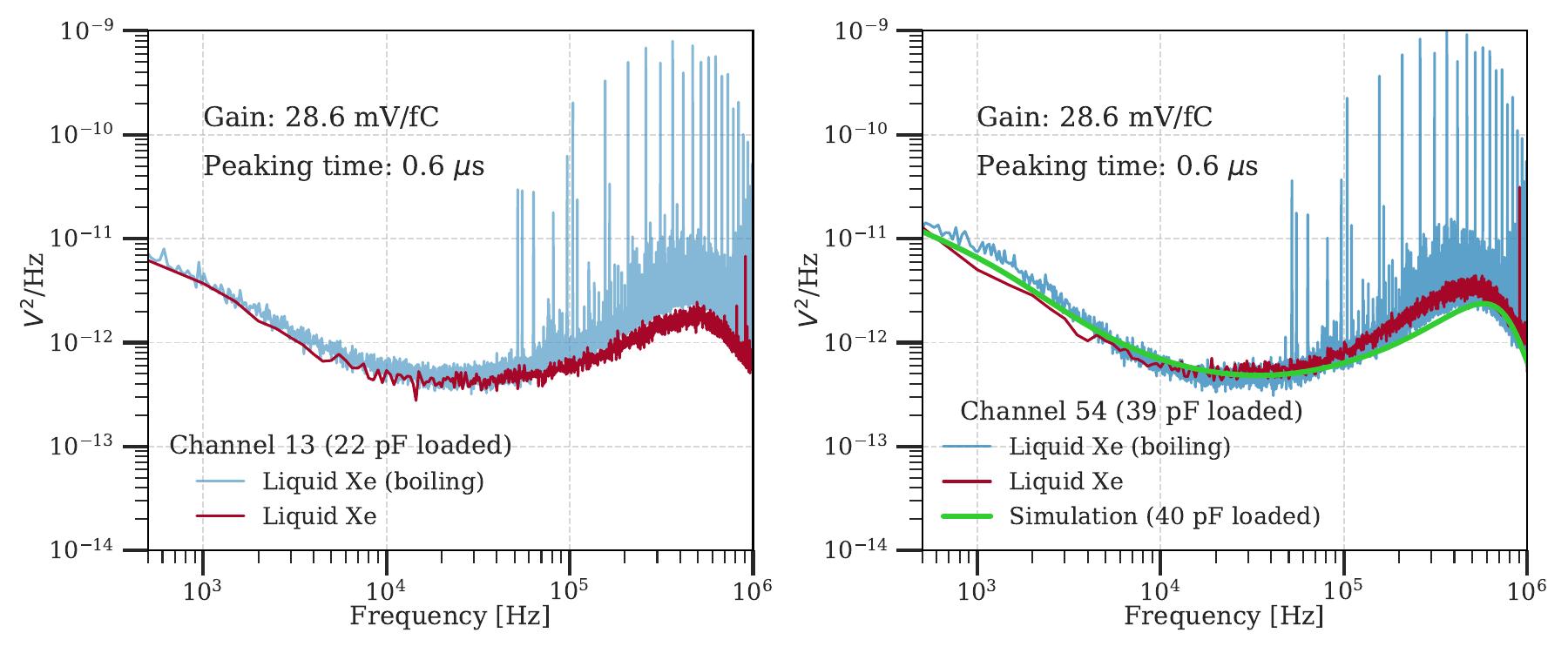}
    \caption{Noise PSD of two representative channels from the top and bottom banks, measured in the LXe test stand, with the front-end channels configured with a 
    gain of 28.6 mV/fC gain and a peaking time of 0.6~$\mu$s. Left: channel 13 from the top bank, loaded with a 22~pF capacitor. Right plot: channel 54 from the bottom bank, loaded with a 39~pF capacitor. The simulated spectrum is shown as the green line and agrees well with the measurements.
    The influence of the liquid boiling issue can be observed on the blue curves. }
    \label{fig:coldboard_LXeTest_PSD}
\end{figure}

For the noise measurements, a total of 100 traces are recorded in sequence, with 4000 ADC samples collected per channel in each trace. The average noise PSD measured at LXe temperature, for two representative channels with 22~pF and 39~pF capacitance loaded, respectively, is shown in Fig.~\ref{fig:coldboard_LXeTest_PSD}, with a gain of 28.6~mV/fC and a peaking time of 0.6~$\mu$s in order to capture the full frequency-domain response. As expected, higher noise power is observed for channels with larger input capacitance. The simulated PSD curve for a 40~pF load is shown as the green line in the right panel, demonstrating good agreement between in-LXe data and design simulations.

\begin{figure}[!h]
    \centering
    \includegraphics[width=0.95\linewidth]{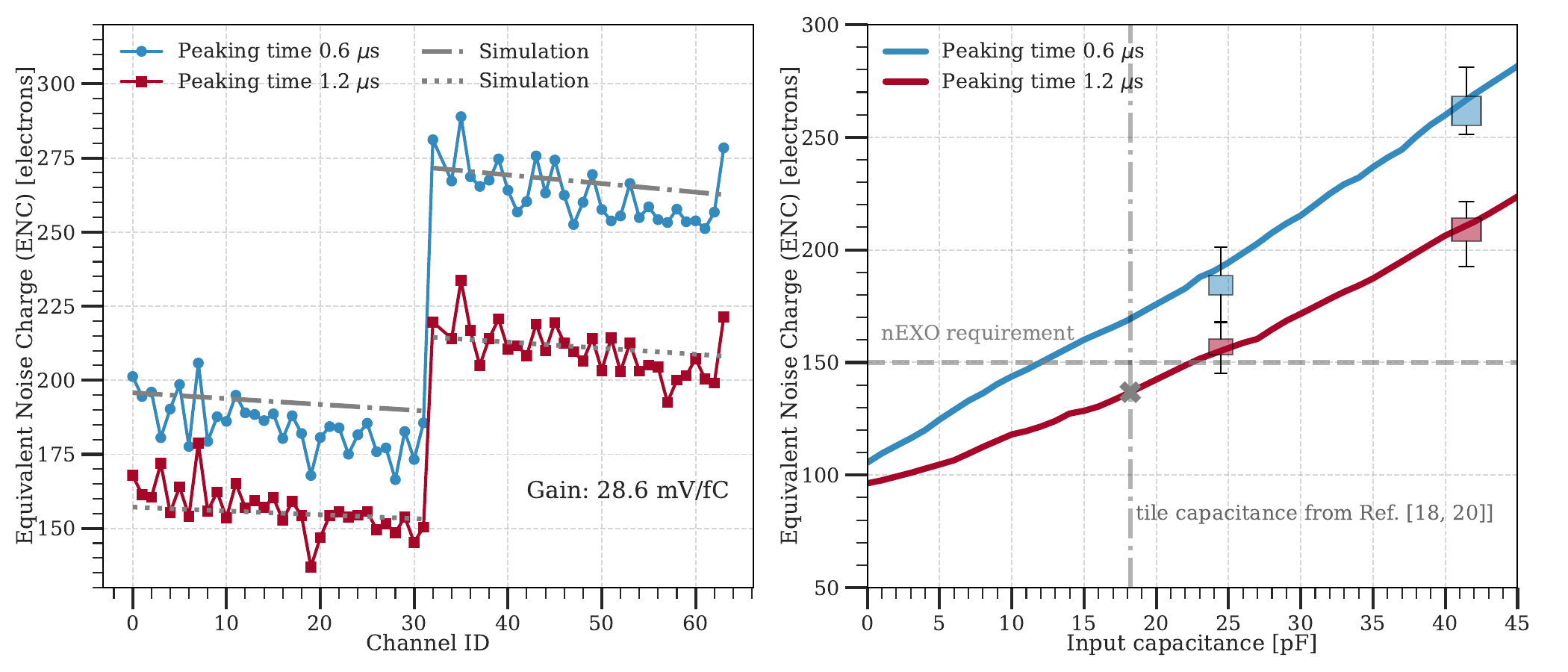}
    \caption{Noise measurements in the LXe test stand. Left: measured noise ENC of all 64 channels on the Cold Board with gain set to 28.6 mV/fC, shown for both 0.6 $\mu$s (red markers) and 1.2 $\mu$s (blue markers) peaking times. 
    Channels 0-31 are loaded with $C_\mathrm{load}$ = 22~pF, while channels 32-63 are loaded with $C_\mathrm{load}$ =39~pF. 
    The gray lines represent the corresponding simulation results. 
    The slight negative slope in the ENC is attributed to the estimated capacitance of the input traces on the Cold Board. 
    Right: Simulated ENC as a function of input capacitance, with the measured ENC values overlaid as box plots. 
    The measurements show good agreement with the simulation. 
    The nEXO noise requirement of 150 $e^-$ is indicated by the gray horizontal line, while the gray vertical line marks the estimated 18.2~pF input capacitance for nEXO. 
    %The agreement between measurement and simulation suggests that the required noise performance can be achieved in nEXO.
    }
    \label{fig:coldboard_LXeTest_noiseENC}
\end{figure}

The standard deviation (STD) of the baseline fluctuations from the ASICs, in units of ADC counts, can be converted to an equivalent noise charge, as shown in the left panel of Fig.~\ref{fig:coldboard_LXeTest_noiseENC}. As expected, the noise level of channels 0–31 (top bank), loaded with 22~pF, is lower than that of channels 32–63 (bottom bank), which are loaded with 39~pF. One non-functional channel (channel 33), likely resulting from damage incurred during testing, has been excluded from the plot. Since nEXO is most likely to read out charge signals with either \SI{0.6}{\mu s} or \SI{1.2}{\mu s} peaking time in order to preserve waveform shape information, we show noise measurements for both configurations. Data taken with a \SI{1.2}{\mu s} peaking time exhibits lower ENC than those with a \SI{0.6}{\mu s} peaking time, due to stronger suppression by the anti-aliasing filter. Channels with the same load and peaking time show a $\sim$10\% variation in noise, which is attributed to differences in wirebonding and trace lengths. By incorporating the parasitic trace capacitance on the Cold Board, simulated noise values across the channels in each bank are shown as gray lines. The small negative slope, which stems from differences in capacitance from the length of each channel's input trace on the PCB, is accounted for by the simulations and agrees well with the measured data. 
For a 22~pF load capacitance, the predicted ENC values are approximately 185~$e^{-}$ for a 0.6~$\mu$s peaking time and 156~$e^{-}$ for a 1.2~$\mu$s peaking time. For a 39~pF load capacitance, the corresponding values from the recent simulation~\cite{CRYOASIC:2023CPAD} are approximately 263~$e^{-}$ and 210~$e^{-}$, respectively. The simulated ENC as a function of input capacitance is shown in the right panel of Fig.~\ref{fig:coldboard_LXeTest_noiseENC}, with the measured data overlaid. The ENC values obtained from the LXe measurements are in good agreement with the simulation. For comparison, we show the calculated single-channel input capacitance for the nEXO charge tile, which is 18.2~pF~\cite{nEXO:2018ylp,Cen:2019ejo}, marked as the gray vertical line in Fig.~\ref{fig:coldboard_LXeTest_noiseENC}. At 18.2~pF, the expected ENC is around 137~$e^{-}$. Therefore, under the configuration with a gain of 28.6~mV/fC and a peaking time of 1.2~$\mu$s, the noise level of CRYO ASIC integrated with a nEXO charge readout tile is expected to satisfy the project's requirement of 150 $e^-$ single channel noise level.

\section{Summary and outlook} 
The performance of CRYO ASIC has been characterized using both the Cold Board and the FEMB board in a cryogenic N$_2$ environmental chamber and a LXe test stand. 
The ASIC functionality is verified at room temperature and at cryogenic temperature relevant to nEXO. In the cryogenic chamber at 165~K, the channel gain remains stable over a 14-hour period, and the on-chip charge injection pulser provides a reliable method for in-situ calibration. During operation in LXe, heat dissipation from the ASIC is initially observed to induce localized xenon boiling, resulting in elevated noise. This effect is mitigated by pressurizing the chamber while minimizing the temperature change, restoring the CRYO ASIC to optimal noise performance. Since nEXO is expected to operate at pressures around 0.15~MPa, these results indicate the ASIC operation will not induce boiling under the planned detector conditions. Under these conditions, the ASIC achieves an average noise performance of 156~$e^{-}$ ENC at a gain of 28.6~mV/fC, a peaking time of 1.2~$\mu$s, and an input capacitance of 22~pF. The measured noise in LXe agrees with design simulations, which indicates that the nEXO charge module can meet the experimental noise requirements. The noise stability should also be cross-checked in a longer-duration measurement in the future.

Future work will involve testing the ASICs coupled to sensor tiles for ionization charge readout, as well as integration tests with the radiopure cables~\cite{Arnquist:2023gtq} and mechanical assemblies required for deployment in the nEXO experiment. The results presented in this paper demonstrate that the CRYO ASIC can achieve low-noise and stable performance when operated directly in LXe, validating its suitability for in-LXe charge readout in nEXO. In addition to the single-chip characterization reported here, comprehensive studies of production yield and large-channel-count quality control will be carried out in later stages of the project.

\section{Acknowledgments}
The authors thank Mark Convery (SLAC) for his significant contributions to the development of the FEMB system, and Bob Conley (SLAC) for his dedicated support in the setup and configuration of the cryogenic testing.
G. R. and A. A. were supported by the U.S. Department of Energy Science Graduate Student Research through contract number DE-SC0014664. 
M.Y. and B.L. are supported by the Department of Energy, Laboratory Directed Research and Development program at SLAC National Accelerator Laboratory, under contract DE-AC02-76SF00515.
This work was also supported in part by Laboratory Directed Research and Development (LDRD) programs at Brookhaven National Laboratory (BNL), Lawrence Livermore National Laboratory (LLNL), Pacific Northwest National Laboratory (PNNL), and SLAC National Accelerator Laboratory. The authors gratefully acknowledge support for nEXO from the Office of Nuclear Physics within DOE’s Office of Science under grants/contracts %DE-AC02-76SF00515, 
DE-FG02-01ER41166, DE-SC002305, DE-FG02-93ER40789, DE-SC0021388, DE-SC0012704, DE-AC52-07NA27344, DE‐SC0017970, DE-AC05-76RL01830, DE-SC0012654, DE-SC0021383, DE-SC0014517, DE-SC0024666, DE-SC0020509, DE-SC0024677 and support by the US National Science Foundation grants NSF PHY-2111213 and NSF PHY-2011948; from NSERC SAPPJ-2022-00021, CFI 39881, FRQNT 2019-NC-255821, and the CFREF Arthur B. McDonald Canadian Astroparticle Physics Research Institute in Canada; from IBS-R016-D1 in South Korea; and from CAS in China.

%\paragraph{Note added.} This is also a good position for notes added
%after the paper has been written.

\bibliographystyle{JHEP}

\bibliography{references}


\end{document}